\documentclass{bmcart}

\usepackage[utf8]{inputenc} 

\usepackage{amsmath}

\usepackage{multirow}
\usepackage{tabulary}
\usepackage{subcaption}
\usepackage{float}
\usepackage{xcolor}
\usepackage{graphicx}
\usepackage{epstopdf}
\usepackage{adjustbox}
\usepackage{graphicx}
\usepackage{endnotes}
\let\footnote=\endnote

\startlocaldefs
\endlocaldefs

\begin{document}

\begin{frontmatter}

\begin{fmbox}
\dochead{Research}


\title{Linking Twitter Events With Stock Market Jitters}


\author[
   addressref={aff1},                   
   corref={aff1},                       
   email={t.tsapeli@cs.bham.ac.uk}   
]{\inits{F}\fnm{Fani} \snm{Tsapeli}}
\author[
   addressref={aff2},                   
   email={nikosbezes@gmail.comk} 
]{\inits{N}\fnm{Nikolaos} \snm{Bezirgiannidis}}
\author[
   addressref={aff1},
   email={P.Tino@cs.bham.ac.uk}
]{\inits{P}\fnm{Peter} \snm{Tino}}
\author[
   addressref={aff3},
   email={m.musolesi@ucl.ac.uk}
]{\inits{M}\fnm{Mirco} \snm{Musolesi}}


\address[id=aff1]{
  \orgname{School of Computer Science, University of Birmingham}, 
  \street{Edgbaston},                     %
  \postcode{B15 2TT}                                
  \city{Birmingham},                              
  \cny{United Kingdom}                                    
}

\address[id=aff2]{
  \orgname{Department of Electrical and Computer Engineering, Democritus University of Thrace}, 
  \street{},                     %
  \postcode{}
  \city{Xanthi},                              
  \cny{Greece}                                    
}
\address[id=aff3]{
  \orgname{Department of Geography, University College London}, 
  \street{Gower Street},                     %
  \postcode{WC1E 6BT}                                
  \city{London},                              
  \cny{United Kingdom}                                    
}


\begin{artnotes}
\end{artnotes}

\end{fmbox}


\begin{abstractbox}

\begin{abstract} 
Predicting investors reactions to financial and political news is important for the early detection of stock market jitters. Evidence from several recent studies suggests that online social media could improve prediction of stock market movements. However, utilizing such information to predict strong stock market fluctuations has not been explored so far.

In this work, we propose a novel event detection method on Twitter, tailored to detect financial and political events that influence a specific stock market. The proposed approach applies a bursty topic detection method on a stream of tweets related to finance or politics followed by a classification process which filters-out events that do not influence the examined stock market. We train our classifier to recognise \emph{real} events by using solely information about stock market volatility, without the need of manual labeling. We model Twitter events as feature vectors that encompass a rich variety of information, such as the geographical distribution of tweets, their polarity, information about their authors as well as information about \emph{bursty} words associated with the event. We show that utilizing only information about tweets polarity, like most previous studies, results in wasting important information. We apply the proposed method on high-frequency intra-day data from the Greek and Spanish stock market and we show that our financial event detector successfully predicts most of the stock market jitters.
\end{abstract}



\end{abstractbox}

\end{frontmatter}

\section{Introduction}
\label{intro}

Predicting investors reactions to financial and political news is important for the early detection of stock market jitters. Evidence from several recent studies suggests that online social media could improve prediction of stock market movements \cite{bollen2011twitter, zhang2011predicting, mao2011predicting}. However, utilizing such information to predict strong stock market fluctuations has not been explored so far.

We propose a novel framework for detecting financial events on Twitter that impact a specific stock market. The proposed financial event detector (FED) monitors the arrival rates of individual words in a stream of tweets and records an event when an unusual burst is detected. For each event we create a feature vector containing information such as the number and type of words with \emph{unusual} increase on their arrival rates, the volume and the polarity of the related tweets as well as geographical characteristics of the tweets and information about their authors. Then, we exploit stock market data in order to train a classifier to recognize as \emph{positive} events that influence stock market and as \emph{negative} events with minimum or no impact. Thus, our method is trained to detect financial or political events that cause fluctuations on a specific stock market. Our method does not require any manual events labeling. Instead, we create a  training set by labeling as \textit{positive} event vectors that co-occur with large movements on the examined stock market and as \textit{negative} all the other vectors. Our classifier is updated dynamically: after a new event (from the remaining set) is classified, the true event label is learned by examining its impact on the stock market and the classifier is re-trained accordingly. 

Typically, social media information is modeled as one-dimensional time-series, describing the evolution of a feature, such as tweets polarity or volume, over time. However, we show here that using a single feature to model Twitter data results in wasting important information. For example, negative criticism about an educational or health system reform could be a popular topic within a country and result in bursts of tweets with negative sentiment; nevertheless it will probably have minimum or no impact on the stock market. On the other hand, news related to financial or political instability would probably be commented by a larger number of different people and may have more global interest. Consequently, features such as the number and the profile of different users discussing a topic, the geographical characteristics of tweets (i.e. whether the topic of interest is discussed locally or globally) as well as the individual \emph{bursty} words associated with each event may contain important information. Thus, instead of arbitrarily selecting a single feature for representing Twitter information, we apply feature selection in order to find an optimal subset of features that can be used to identify vectors reflecting important information about the examined stock market. We support our argument by comparing the proposed FED method with a modified version that uses single-feature vectors, which contain only information about events sentiment. 
 
We apply the proposed framework on the detection of events that influence the Greek and Spanish stock market for the period of 8 months. More specifically, we collect Twitter data by tracking terms related to the European crisis and we examine the impact of detected events on intraday returns of the ATHEX and IBEX stock market indexes. We select these two markets given the fact that several events that influenced them took place recently. Moreover, the number of truly global companies quoted in them is fairly limited and, therefore, it is possible to identify and assess national events that might affect them. We show that the proposed method achieves up to 74\% and 68\% F1 scores on the detection of Twitter events that influence ATHEX and IBEX indexes respectively. Moreover, we apply a general-purpose state-of-the-art event detector on our Twitter dataset and we demonstrate that such approaches fail to recognize events which influence stock market.
It is also worth noting that one of our key design goals is to minimize the use of thresholds. In fact, we are aware that the choice of their values might be difficult and, for this reason, our aim is to propose a solution that relies on a minimal number of them. We discuss criteria for the choice of the values of the thresholds, also considering their impact on the overall performance of our approach.

The rest of this paper is organized as follows. In Section \ref{relatedWork} we discuss related work on both event detection on Twitter and on its influence on the stock market. In Section \ref{eventDetectionTwitter} we present the proposed financial event detector and in Section \ref{applicationGreece} we demonstrate the performance of the method on the ATHEX and IBEX stock market indexes in comparison with a general-purpose event detector and a sentiment-based financial event detector. Additionally, we demonstrate the dependence between Twitter events and stock market jitters by conducting a mutual information analysis. Finally, in Section \ref{conclussions} we summarize the findings of our work.

\section{Related Work}
\label{relatedWork}

\subsection{Event Detection on Twitter}

Detecting emerging topics with wide interest on Twitter has recently gained the attention of researchers and practitioners. The main approach that is followed is the detection of \emph{bursty} terms (i.e., words or segments that their frequency on the twitter stream experiences some unusual pattern during a specific time period) followed by a grouping of these terms based on their content similarity or similarities on their arrival patterns. For example, in~\cite{cataldi2010emerging} the authors extract a set of \emph{emerging} terms from the twitter stream by assigning weights to each term based on its frequency as well as the \emph{importance} of twitter users who use it and keeping only the highest weighted terms. The \emph{emerging} terms are grouped together by examining their co-occurrence on the same tweets. EDCoW~\cite{weng2011event} constitutes also a representative example of this category of event detectors. EDCoW performs a wavelet transformation on the frequencies of words and it uses \emph{vanishing} auto-correlations to eliminate words that do not experience any irregularities on their arrival rates. Afterwords, it creates a graph by using the pairwise correlations among the words wavelets and applies a modularity-based graph partitioning in order to group the words to events. On the same direction, TopicSketch \cite{xie2013topicsketch} monitors the acceleration of words and pairs of words in order to early detect bursty topics. Twevent~\cite{li2012twevent} proposes the use of word segments instead of single words and detects bursty words by examining the word segments frequency and the number of different users that report these segments. Then, word segments are grouped by examining the content similarity among them. 

Other projects focus on detecting events of a specific type. For example, authors in~\cite{lee2010measuring} detect local social events by monitoring microblogging activity in geographical regions and reporting any unusual activity. Moreover, in~\cite{sakaki2010earthquake} a system for detecting real world events in real-time along with the geographical location of the event is presented. The system uses keywords to detect specific event types. Authors present a case study for earthquakes detection. 

Our work substantially differs from the existing event detectors, since our objective is the detection of events that influence a specific stock market rather than the detection of events in general or events of a specific type (e.g., financial events). Thus, our system does not consider real world financial or political incidents that do not impact the examined stock market as \emph{true} events.
 
\subsection{Social Media Influence On Stock Market}

Recently, several studies have shown that information posted on social media or, in general, from web-related sources such as search engines volumes, influences stock market prices~\cite{preis2010complex, mao2011predicting, preis2013quantifying}. Most studies extract a sentiment index reflecting the general pessimism or optimism of tweets and they demonstrate the relationship of this index with the prices of trading assets by conducting correlation analysis~\cite{zhang2011predicting, ruiz2012correlating}, Granger causality analysis~\cite{bollen2011twitter} or multivariate regression analysis\cite{mao2011predicting, deng2011combining, schumaker2009textual}. Moreover, in~\cite{vu2012experiment} the authors train a classifier to predict daily up and down movements of tech companies traded assets using as features the sentiment of relevant tweets and the degree of stock market confidence. 

All of the above-mentioned studies attempt either to predict stock market movements or prove a causal link between stock market and social media. Here, we tackle a different problem: we use social media in order to detect financial/political events that influence stock market. We also show that tweets polarity, which has mainly been used by previous studies, does not fully reflect all the significant information of social media.

\section{Financial Event Detector}
\label{eventDetectionTwitter}
In this section we describe the components of our financial event detector (FED). Our event detection process is comprised by the following steps:
\begin{enumerate}
\item \textbf{Bursty Words Detection.} The arrival rate of each word in a stream of tweets is estimated and a set of \textit{bursty} words is extracted.
\item \textbf{Events Feature Vectors Extraction.} The busty words as well as information extracted from tweets containing these bursty words (such as information related to tweets polarity, geographical distribution and users characteristics) are used to create feature vectors that represent events. 
\item \textbf{Events Filtering.} All the detected Twitter events are not necessarily related to stock market jitters. We use stock market data to train a classifier to recognize which event feature vectors do have an impact on the stock market. Our financial event detector is trained for a specific stock market. The initial labeled training set is created by utilizing solely stock market data, without the need of manual labeling, and the classifier is updated dynamically. 
\end{enumerate}

\subsection{Bursty Words Detection}
\label{burstyWords}
In order to detect bursty topics on a stream of tweets we apply a feature-based event detection method, according to which the arrival rate of each word/feature contained in each tweet is modeled as an inhomogeneous Poisson process. Let us denote by $N_w$ the number of occurrences of each word $w$ in the collection of tweets. We estimate the arrival rate $\lambda_w(t)$ of word $w$ as follows:

\begin{equation}
	\lambda_w(t) = \sum_{i=1}^N {f_{\Delta}(t-t_i)}
\end{equation}
where $t_i$ the time that the $i^{th}$ tweet containing the word $w$ was posted and $f_{\Delta}$ a Gaussian kernel of bandwidth $\Delta$. We characterize a word $w$ as \emph{bursty} during a specific time interval by applying thresholds both on the rate of the word $\lambda_w(t)$ and on the slope $\lambda_w'(t)$ of its rate.  In detail, a word $w$ which was not \emph{bursty} at time $t-1$ will be \emph{bursty} at time $t$  if $\lambda_w(t) > T_R$ and $\lambda_w'(t) > T_S$, for some threshold values $T_R, T_S>0$. A word $w$ \emph{bursty} at time $t$ will not be \emph{bursty} at time $t'>t$ if $\lambda_w(t') < T_R$. Hence, we examine only the rate of the word in order to change its status from \emph{bursty} to \emph{normal} since, even if the acceleration of the rate of a previously characterized \emph{bursty} word is low, the word should still be considered as \emph{bursty} if it has a sufficiently high arrival rate. The rate of each word is re-computed dynamically every time that a tweet containing that word is posted. 

By applying a threshold on the word rates, we detect words with significant \emph{popularity} (i.e. high rate) within a time-period, while by applying a threshold on the rate slope we avoid considering as \emph{bursty} words which are \emph{popular} most of the time. We use the same thresholds for all the words instead of creating word-specific thresholds based on historical rates. In this way, we not only reduce the number of free parameters but also avoid considering as \emph{bursty} words which occur a relatively small number of times during a period while they previously had zero or very low rate.

\subsection{Events Feature Vectors Extraction}
\label{eventDetection}

An event on Twitter at time $t$ will exist if there is at least one bursty word at time $t$. We represent events as time-dependent feature vectors (a detailed description of the features is presented at section \ref{featuresDescription}). Only one event feature vector can be active during any time $t$. Hence, if more than one Twitter events co-occur, they will all be associated with the same feature vector and all bursty words describing these events will comprise different features of the vector. An event starts when a bursty word is detected and it is updated dynamically every time a \emph{significant} change on its characteristics occurs. Thus, multiple feature vectors may be created for the same event. In order to avoid unnecessary overhead we do not update an event every time an insignificant change occurs in any of the feature values. Instead, we consider that a significant change on the over-all arrival rate of all the words that are associated with the event indicates a noteworthy change on the volume and characteristics of tweets and therefore, the event features need to be re-estimated. We consider an increase in the overall arrival rate of words to be significant if it is at least 10\% of the previous value. Thus, if $W$ the set of bursty words associated with an event, a new feature vector for the same event that had its last update at time $t_1$ will be created at time $t_2$ if:

\begin{equation}
	\sum_{w \in W}{\lambda_w(t_2)} > 1.1 \cdot \sum_{w \in W}{\lambda_w(t_1)} 
\label{eq:eventUpdateCond}
\end{equation}

The feature vector that corresponds to the most recent event update is created so that it represents the \emph{strongest} version of the event. More formally, denote by: $E_{t_s}(t)$ the event started at time $t_s$ and updated at time $t$, $f_i(t)$ the $i^{th}$ feature of the feature vector $E_{t_s}(t)$, $N$ the number of features, $T_u$ the set of the timestamps at which the event $E_{t_s}$ has been updated and $t_l \in T_u$ the last time that the event has been updated. Then, the feature vector $E_{t_s}(t_l)$ is estimated as follows:

\begin{equation}
	E_{t_s}(t_l) = \{\underset{t \in T_u}{\max} f_1(t), \underset{ t \in T_u}{\max} f_2(t), ...\underset{ t \in T_u}{\max} f_N(t)\}
\label{eq:updatedEvent}
\end{equation}

The rational behind this idea is that events need to be detected as early as possible, thus the conditions for creating an event should be relatively 'soft'. However, an initially \emph{weak} event may become stronger later and consequently the initial feature vector will not fully represent the strength of the event. On the other hand, if an event occurs when the stock market is closed, its strength may decrease by the time the stock market opens. Nevertheless, a reaction of the stock market to the news is still expected. Thus, we decided to update an event only when its \emph{strength} has increased. The event will be considered \emph{inactive} when all the words associated with it are not bursty any more. In order to avoid storing obsolete information in the feature vectors, we do not allow events to be active for more than 24 hours. 

\begin{figure}[!th]
  \centering
\includegraphics[width=0.7\textwidth]{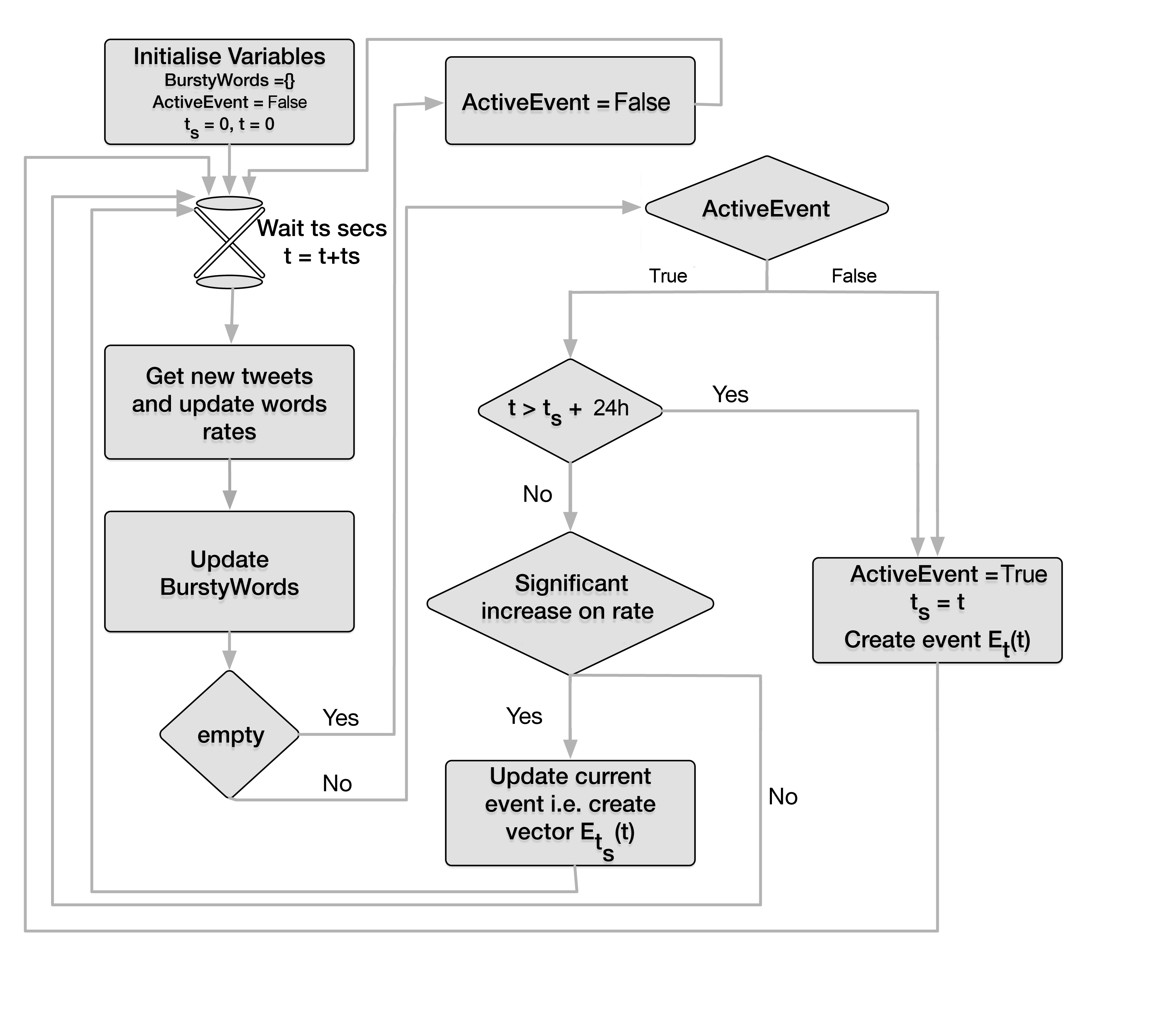}
\caption{Event Detection Process.}
\label{fig:eventsDetection}
\end{figure}

The process of event detection and update is summarized in Figure \ref{fig:eventsDetection}. In detail, we check for new tweets every $\Delta t$ seconds and we update the rate functions of all the words that are contained in the new tweets. Then, the current set of bursty words is updated accordingly (i.e. words which are not bursty any more are removed from the set and new words which are bursty at the current time are added). While the set of bursty words is empty, we check for new tweets every $\Delta t$ seconds until at least one word contained in the tweets becomes bursty. Then, if there is no active event (i.e. the set of bursty words was empty $\Delta t$ seconds before) we create a new event (i.e. a new feature vector $E_t(t)$, where $t$ the current time). If there is an active event and 24 hours have elapsed since its start time, the current event is inactivated and a new event is created; otherwise, we create a new feature vector, as described in equation \ref{eq:updatedEvent}, if the condition of equation \ref{eq:eventUpdateCond} is met. 

\subsubsection{Features Description}
\label{featuresDescription}

The feature vectors include information about the bursty words that are associated with the event, the tweets polarity and geographical distribution and the reputation and popularity of tweets authors. Instead of using a separate feature for each bursty word, we create categories of words that usually refer to the same subject by estimating the correlation between the words rates. Words with highly correlated rates (i.e. similar arrival patterns) may refer to the same subject \cite{weng2011event}. We group words by performing hierarchical clustering, where the 'distance' between two words $w_1$, $w_2$ with correlation coefficient $c_{w_1, w_2}$ is equal to $1 - c_{w_1, w_2}$. The event $E_{t_s}(t)$ started at time $t_s$ will be described by the following features at time $t$:

\begin{itemize}
	\item the \emph{maximum number of bursty words} of each category of words that have been bursty from the start of the event $t_s$ until time $t$. We denote by $W_i(t)$ the set of bursty words of the $i^{th}$ category from time $t_s$ until time $t$ and with $|W_i(t)|$ the number of elements in the set. Thus, there is one feature $|W_i(t)|$ for each category $i$ of words. 
	\item the \emph{maximum bursty word rate in each word category $i$:}  $R_i(t) = \underset{w \in W_i(t)}{\max} \underset{t_s \leq t' \leq t}{\max} \lambda_w(t')$ 
	\item the \emph{maximum bursty word rate:}  $R(t) = \underset{i \leq N}{\max}R_i(t)$.
	\item the \emph{maximum bursty word rate slope:} \\$S(t) =  \underset{i \leq N}{\max}\underset{w \in W_i(t)}{\max} \underset{t_s \leq t' \leq t}{\max} \frac{d}{dt'}\lambda_w(t')$.
	\item the \emph{number of verified users} $V(t)$ that have posted a tweet which contains at least one bursty word from the start of the event until time $t$, normalized by the number of tweets \footnote{Note that a verified user will be counted for as many tweets as she will post in the event time interval.}. Verified accounts are, according to Twitter, highly sought users in interest areas including government, politics, journalism, business, etc., and thus include authenticated accounts of the key players in major political and economic events.
	\item the \emph{average number of followers} $F_{AVG}(t)$ of users who have posted a tweet containing at least one bursty word. The number of followers of a Twitter user is an indication of the impact his/her tweets may have.
	\item the \emph{maximum number of followers} $F_{MAX}(t)$ between all the users who have posted a tweet which contains at least one bursty word. 
	\item the \emph{average geographical distance from the examined stock market location} $D_{AVG}$ of users who have posted a tweet associated with the event. 
	\item the \emph{weighted average distance from the examined stock market location} $D_{W\_AVG}$: calculated as $D_{AVG}$, but this time each user is weighted by the corresponding proportion of the followers, i.e. the number of her followers normalized by the total number of all users followers. 
	\item the \emph{location dispersion} $L(t)$ of the users who have posted a tweet associated with the event. The location dispersion is an indication of whether the topic is discussed mainly locally or whether there is a general interest for the event globally. It is calculated using the coefficient of variation (i.e. the ratio of the standard deviation to the mean) of the distance from the event centre, among all event tweets.
	\item the \emph{sentiment strength index} $SSI(t)$. We use SentiStrength \cite{thelwall2013heart} in order to estimate the positivity and negativity index of the tweets that are associated with the event, and calculate the tweet sentiment strength index as the sum of these two indices. The event sentiment strength index is calculated as the absolute value of the average sentiment strength index among all event tweets. To emphasize the \emph{financial sentiment} of the tweets, we have trained the SentiStrength sentiment index calculation by manually optimizing the term strengths using high-financial-impact words.
	\item the \emph{weighted sentiment strength index} $SSI_W(t)$, which is the average sentiment strength index among all tweets of the event, weighted by the number of followers of each tweet author (as in $D_{W\_AVG}$).
\end{itemize}

Thus, if $N$ the number of word categories, an event will be characterized by a feature vector of $2 \cdot N + 10$ features: $E_{t_s}(t)=$ $\{|W_1(t)|$, $R_1(t)$, $|W_2(t)|$, $R_2(t)$, ... $|W_N(t)|$, $R_N(t)$, $R(t)$, $S(t)$, $V(t)$, $F_{AVG}(t)$, $F_{MAX}(t)$, $D_{AVG}$, $D_{W\_AVG}$, $L(t)$, $SSI(t)$, $SSI_W(t)$\}. All features are normalized to zero mean and unit variance. As it is described in the following section, feature selection will be applied in order to distinguish the features which are actually important for the detection of events which influence the examined stock market.

\subsection{Event Filtering}
\label{eventsFiltering}

The last step of our event detection process is filtering out events that do not have any impact on the examined stock market. In order to classify the events to \textit{positive} (i.e. events that have an impact on stock market) and \textit{negative} (i.e. events that do not influence stock market), we train a classifier by utilizing stock market data to create a labeled training set. Thus, the classifier is trained to recognize events that influence a specific stock market. In detail, we first define a set of time-slots $\mathcal{T}_{true}$ during which an \textit{unusual} movement on stock market volatility is noticed and a set of time-slots $\mathcal{T}_{false}$ during which the volatility is considered \textit{normal}. Let us denote with $\mathcal{V}(s)$ the volatility of stock market at time $s$ and $\mathcal{V'}(s)$ the volatility slope. Then, we set $s \in \mathcal{T}_{true}$ if $\mathcal{V'}(s)>T_{true}$ and $s \in \mathcal{T}_{false}$ if $\mathcal{V'}(s)<T_{false}$, where $T_{true}>0$ and $T_{false}>0$ are thresholds on the stock market volatility slope. We set $T_{true} > T_{false}$ in order to allow for a \textit{neutral} zone and separate high volatility  time-slots belonging to $\mathcal{T}_{true}$ from the normal volatility ones belonging to $\mathcal{T}_{false}$. We also define with $\mathcal{T}_{neutral}$ the set of time-slots that belong to the neutral zone.

\begin{figure}[!th]
  \centering
\includegraphics[width=0.5\textwidth]{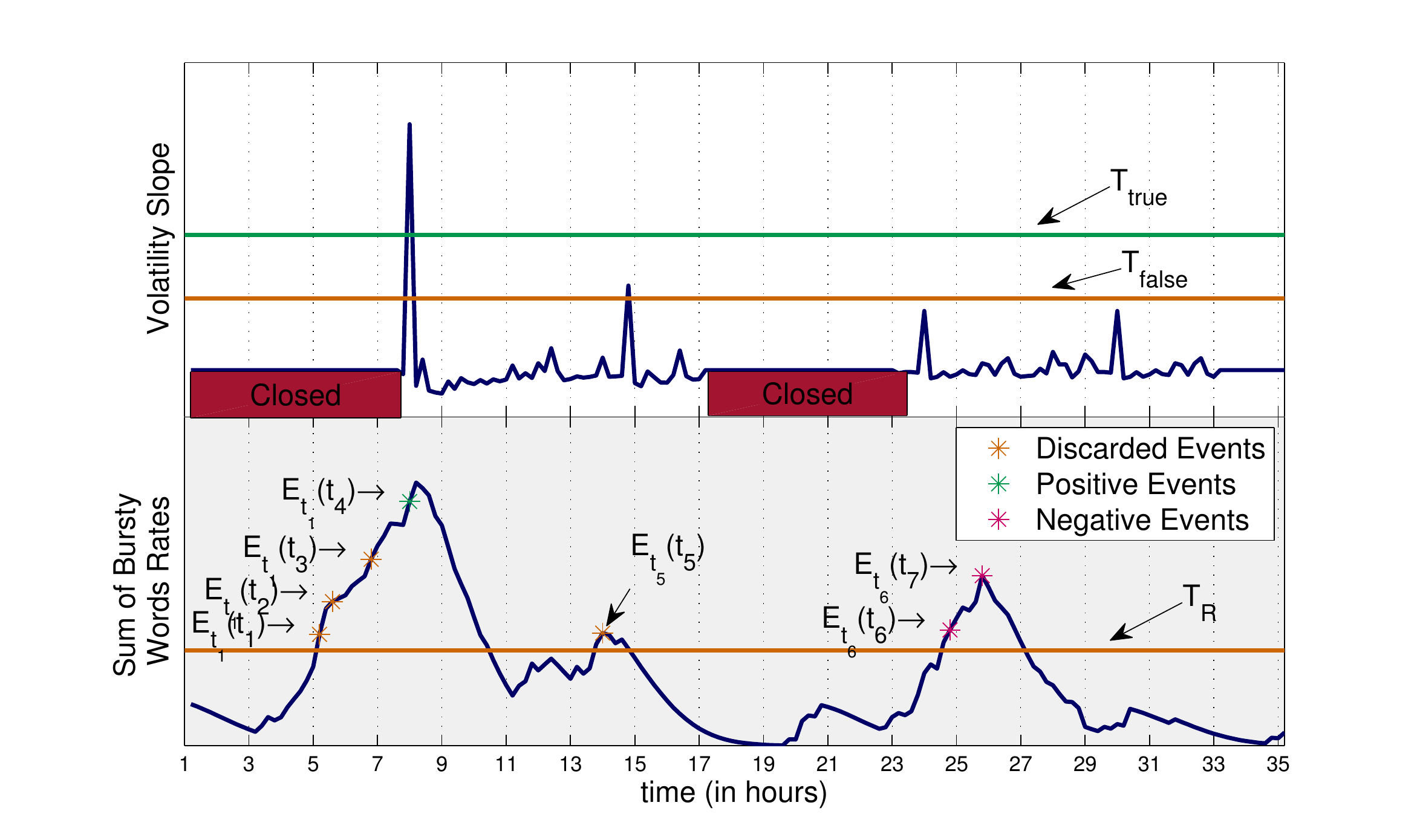}
\caption{Events Labeling Example.}
\label{fig:examplePlot}
\end{figure}

Afterwords, we need to examine which event feature vectors co-occur with unusual movements on stock market volatility. Since a stock market is open only at specific hours, some events on Twitter may occur when the stock market is closed. However, the effect of these events (if any) will be visible only when the stock market opens. In order to match the time at which an event occurred or was updated with the time that its impact was visible at the stock market, we transform the update time $t$ of each feature vector to the first time $t'\geq t$ that the stock market is open. If multiple update times of the same event match with the same stock market time $t'$, we keep only the most recent feature vector and discard all the previous ones; the rational behind this decision is that the most recent feature vector represents the event on its full strength. If multiple events (i.e. events with different start times) match with the same stock market time, we keep all events. Finally, an event that started at time $t_s$, matched to the stock market time $t'$, will be assigned a label $C_{t_s}(t')$ as follows:

\begin{equation}
C_{t_s}(t') = 
\left\{
  \begin{array}{rcr}
	1 & \mbox{if $\underset{s \in \mathcal{T}_{true}}{\min} |s-t'|<T_{time}$}\\
	-1 & \mbox{if $\underset{s \in \mathcal{T}_{neutral}}{\min} |s-t'|<T_{time}$}\\
    0 & \mbox{otherwise}\\
    
\end{array}
\right.
\label{eq:labelVectors}
\end{equation}
where $T_{time}$ a threshold that denotes the maximum time distance that a twitter event may have from a stock market event, value $1$ is used to label \emph{positive} events and value $0$ \emph{negative}. Event feature vectors with label $-1$ will be discarded from the training set. In Figure \ref{fig:examplePlot} we present an events labeling example. The upper graph depicts the volatility slope and the lower the sum of the bursty words rates. In this example, we set $T_{time} = 1$ hour. Event $E_{t_1}$ is detected when stock market is closed and it is updated at times $t_2$, $t_3$ and $t_4$. We discard the first three event vectors and we set the label of the most recent event update (i.e. $E_{t_1}(t_4)$) equal to $1$. Event $E_{t_5}$ will also be discarded since, according to our graph, there is a time sample $s \in \mathcal{T}_{neutral}$ with less than one hour difference from the event detection time $t_5$. Finally, the event vectors $E_{t_6}(t_6)$ and $E_{t_6}(t_7)$ do not co-occur with any stock market jitter so they will be labeled as negative. 

\begin{figure}[!th]
  \centering
\includegraphics[width=0.45\textwidth]{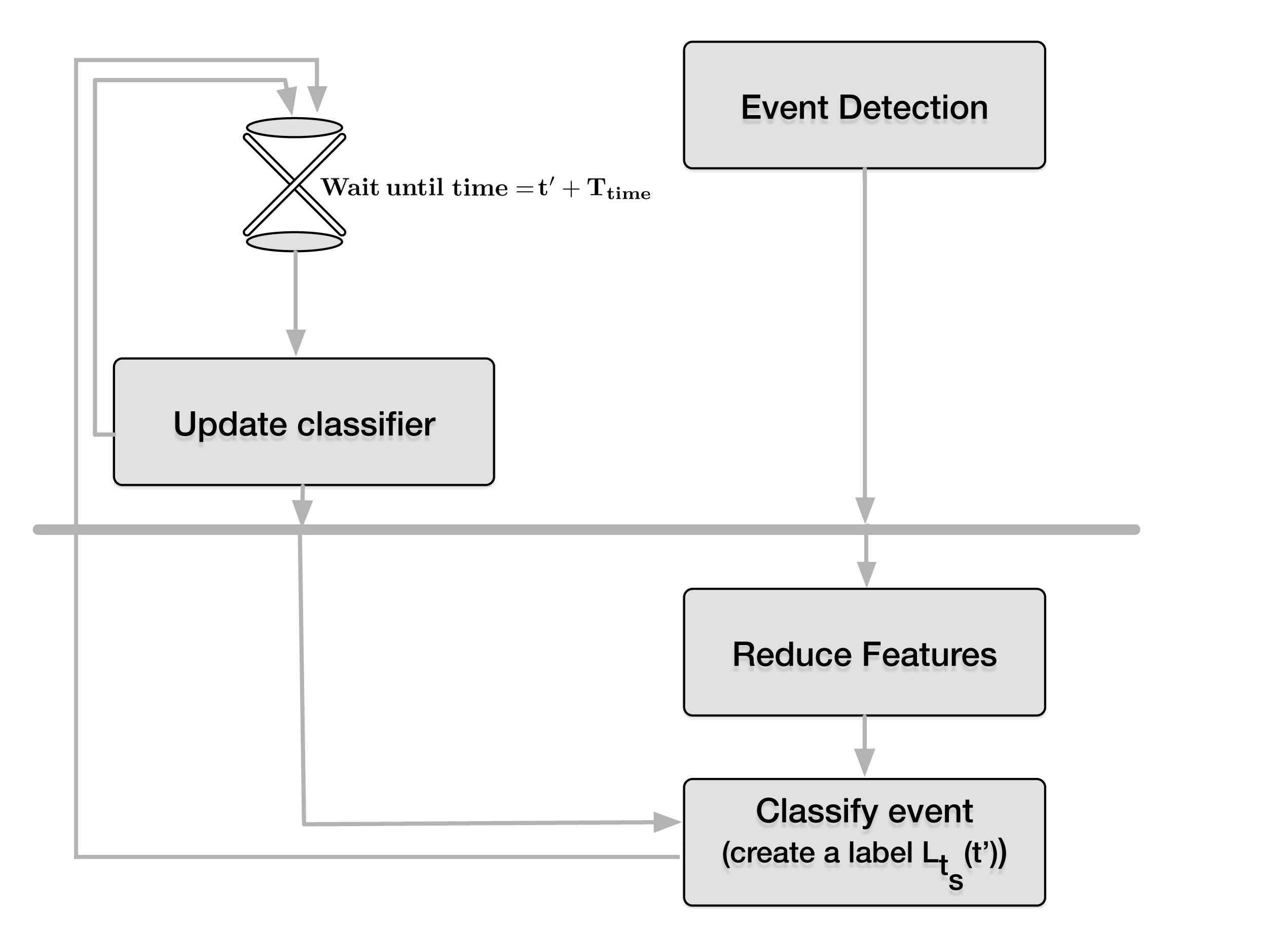}
\caption{Event Classification Process.}
\label{fig:eventsFiltering}
\end{figure}

We create a labeled training set by using a subset of the available data. We reduce the feature set by applying a feature selection process on the training set, thus keeping only features useful for distinguishing \textit{positive} events from \textit{negative} events. Finally, we use these reduced-dimensionality feature vectors to train a classifier. 

The event filtering is performed dynamically. This process is presented at Figure \ref{fig:eventsFiltering}. Each time a new feature vector is created (by the process described in Figure \ref{fig:eventsDetection}), we firstly reduce the vector dimension by discarding insignificant features, we then match to the time $t'$ of the next stock market sample and, finally, we use the previously trained classifier to assign a label $L_{t_s}(t')$ to the event. 

We are able to determine the actual label $C_{t_s}(t')$ (Equation \ref{eq:labelVectors}) of the event only at time $t' + T_{time}$. After the true label $C_{t_s}(t')$ becomes available, we dynamically update the classifier based on the estimated and true class labels $L_{t_s}(t')$ and $C_{t_s}(t')$, respectively.

\section{Application to Stock Markets}
\label{applicationGreece}
We apply the proposed framework to verify whether there is a link between detected events in social media (in our case Twitter) and events (large fluctuations) on the Greek and Spanish stock markets.  In particular, we estimated the historical volatility of the ATHEX and IBEX stock market index by using 5-minutes intraday data for the period 01/03/2015 to 01/11/2015\footnote{Note that the Greek dataset corresponds to a period of around 7 months given that Greek stock market was closed during July}. We also downloaded Twitter data by tracking terms related to the European financial crisis. In order to select appropriate terms, we applied the RAKE keyword extraction method \cite{rose2010automatic} on the European debt crisis Wikipedia webpage\footnote{https://en.wikipedia.org/wiki/European\_debt\_crisis}. We set the maximum number of words per keyword equal to 2 and the minimum number of occurrences in the document equal to 4. 210 keywords were extracted in total. We selected only the keywords with score larger than 1.2 which resulted in 158 keywords in total. The selected keywords include both general terms related to finance such as bank, crisis or debt and other more specific terms related to Europe such as EFSF, Eurozone, European Central bank and so on. The Greek and Spanish Twitter datasets will include only tweets which contain the terms Greece or Greek and Spain or Spanish respectively. We use the data of the first 4 months to find the optimal parameters for our classifier and the remaining data for the evaluation of FED performance. 

\subsection{Thresholds Selection}
One of our key design goals is to minimize the use of thresholds, and in any case to understand and quantify the impact of the choice of different values on the performance of the proposed approach. Thus, in this section we discuss several criteria for the selection of threshold values.
\textbf{Thresholds related to Words Arrival Rate, $\mathbf{T_R, T_S}$.}
The values of these thresholds regulate how often an event is created or updated based on the process described in Figure \ref{fig:eventsDetection}. Large threshold values would result in missing events or in delayed event detection. On the other hand, smaller values would result in more \textit{false positive} events. However, as described in Section \ref{eventsFiltering}, a classifier will be used to filter out false positives.
Thus, we assign relatively small values to thresholds $T_R$ and $T_S$. In detail, we set 
\[
T_R = \underset{w}{\max} \langle\lambda_w(t)\rangle_t \ \ \  \hbox{and} \ \ \ T_S = \underset{w}{\max} \langle\lambda'_w(t)\rangle_t,
\] 
where $\langle\lambda_w(t)\rangle_t$ and $\langle\lambda_w(t)'\rangle_t$ denote the average values of the functions $\lambda_w(t)$, $\lambda'_w(t)$, corresponding to the word $w$. The calculation of the maximum value is over all words in the training data. We observe that for words with an unusual increase in their rate, this mostly happens over short time intervals - most other words have just a constant and very low arrival rate. Hence the resulting thresholds $T_R, T_S$ are very small. 

\textbf{Thresholds on Volatility Slope, $\mathbf{T_{true}, T_{false}}$.} 
The thresholds have to be chosen according to the specific application requirements, i.e., the ``intensity'' of the event under consideration. We examine the performance of FED for three different threshold values. In detail, we set $T_{true} = \{2 \cdot \langle\mathcal{V'}(s)\rangle_s, 2.5 \cdot \langle\mathcal{V'}(s)\rangle_s, 3 \cdot \langle\mathcal{V'}(s)\rangle_s\}$ and $T_{false} = 0.8 \ T_{true}$. 

\textbf{Threshold $\mathbf{T_{time}}$.}
 This threshold denotes the maximum time difference between a Twitter event detection (or update) and a stock market jitter. As mentioned in Section \ref{eventsFiltering}, if the stock market is closed at the time of the event detection, the event detection is formally shifted to the closest opening time of the stock market.
Given that the reaction of stock markets on news is usually instantaneous we set ${T_{time}}$ equal to 1 hour.

\subsection{Feature Selection}

We use the training data to cluster words into categories, as described in subsection \ref{featuresDescription}. We apply hierarchical clustering with cut-off distance equal to 0.7, leading to 58 and 79 word categories for the Greek and Spanish datasets respectively. However, only 34 and 41 of these categories contained at least one bursty word respectively for the two datasets. Thus, since we create two features per word category (i.e. number of bursty words in the category and maximum arrival rate among all words in the category), the total number of word-related features for the Greek dataset is 68 and for the Spanish 82. As described in \ref{featuresDescription}, we also use 10 additional features. To select only features deemed important for distinguishing between \textit{positive} and \textit{negative} events we apply two feature selection methods implemented at Weka \cite{witten1999weka}, namely a correlation-based feature selection algorithm \cite{hall1997feature} and an information gain based feature selection method \cite{yang1997comparative}. Overall, 5 non-word features were selected for the Greek dataset and 6 for the Spanish with both algorithms. The selected attributes along with their ranking based on the correlation-based feature selection are presented at Table \ref{tab:nonWordFeatures} (the results with the information gain based algorithm are very similar and thus they are omitted). Finally, word features related to 7 and 6 word categories were selected, respectively for the two datasets. in Table \ref{tab:featuresSelection}  we present the stemmed words of the selected categories along with the selected features per category. 

\begin{table}[]
\begin{small}
\centering
\begin{tabular}{|c|c|c|}
	\hline
 \multirow{2}{*}{\textbf{Feature}} & \multicolumn{2}{c|}{\textbf{Ranking}}\\
 \cline{2-3}
 & \textbf{GR}& \textbf{ES} \\
  \hline
  \multicolumn{1}{|l|}{maximum rate value $R(t)$ }& 1 & 1\\
   \hline
  \multicolumn{1}{|l|}{maximum slope value $S(t)$} & 2 & 2\\
   \hline
    \multicolumn{1}{|l|}{maximum number of followers $F_{MAX}(t)$} & 3 & 4\\
   \hline
    \multicolumn{1}{|l|}{weighted average distance from stock market location $D_{W\_AVG}$} & 4 & 3\\
   \hline
    \multicolumn{1}{|l|}{weighted sentiment strength index $SSI_W(t)$ }& 5 & 6\\
	\hline
	\multicolumn{1}{|l|}{location dispersion $L(t)$}& - & 5\\
 \hline

\end{tabular}
\caption{Selected non-word features}
\label{tab:nonWordFeatures}
\end{small}
\end{table}

\begin{table}[]
\begin{small}
\centering
\begin{tabular}{|c|>{\centering}m{9cm}|c|c|}
	\hline
 \multirow{2}{*}{} & \multirow{2}{*}{\textbf{Stemmed Words}}&  \multicolumn{2}{c|}{\textbf{Features}}\\
 \cline{3-4}
 &{}& $\mathbf{R_i(t)} $& {$\mathbf{\|W_i(t)\|} $} \\
 	\hline
 1 &  \multicolumn{1}{m{8cm}|}{energy, sovereign, pipelin, sanct} & Yes & Yes\\
 \hline
 2 &  \multicolumn{1}{m{8cm}|}{send, reform, troik} & Yes & No\\
 \hline
 3 &  \multicolumn{1}{m{8cm}|}{money, fear, imf, stock, deb, default, pay, repay} & Yes & Yes\\
 \hline
 4 & \multicolumn{1}{m{8cm}|}{progress, dijsselbloem, finmin, varoufak, min, eurogroup} & Yes & Yes\\
 \hline
 5 & \multicolumn{1}{m{8cm}|}{press, europ, program, fin, govt, bank, deal, stat, nee, bailout, cris, ecb, let, syriz, country, credit, support, eurozon, meet, grexit, econom, polit, euro, greek, agr, talk }& Yes & Yes\\
 \hline
 6 &  \multicolumn{1}{m{8cm}|}{bahrain, eu, leav}& No & Yes\\
 \hline
 7 &  \multicolumn{1}{m{8cm}|}{bil, rep, loan, germ, germany}& No & Yes\\
 \hline

\end{tabular}
\subcaption{Greece}

\begin{tabular}{|c|>{\centering}m{9cm}|c|c|}
	\hline
 \multirow{2}{*}{} & \multirow{2}{*}{\textbf{Stemmed Words}}&  \multicolumn{2}{c|}{\textbf{Features}}\\
 \cline{3-4}
 &{}& $\mathbf{R_i(t)} $& {$\mathbf{\|W_i(t)\|} $} \\
 	\hline
 1 &  \multicolumn{1}{m{9cm}|}{crit, infl, tax, issu} & Yes & Yes\\
 \hline
 2 &  \multicolumn{1}{m{9cm}|}{govern, ban, black, prep, tsipra, money, england, impact, throughout} & Yes & Yes\\
 \hline
 3 &  \multicolumn{1}{m{9cm}|}{mean, chin, ev, europ} & Yes & Yes\\
 \hline
 4 & \multicolumn{1}{m{9cm}|}{neg, greec, inform, comp, demand, spend} & Yes & Yes\\
 \hline
 5 & \multicolumn{1}{m{9cm}|}{reform }& Yes & No\\
 \hline
  6 & \multicolumn{1}{m{9cm}|}{form, elect, perc, ecb, contribut, podemo, rajoy} & Yes & Yes\\
 \hline

\end{tabular}
\subcaption{Spain}
\end{small}
\caption{Selected Word-related Features.}
\label{tab:featuresSelection}

\end{table}

\subsection{Evaluation}

\label{evaluation}

We apply the methodology discussed in Section \ref{eventsFiltering} to train a support vector machine classifier in an online manner to classify the future Twitter events into \textit{positive} and \textit{negative}. 
The classifier is first trained on an initial data segment (training set), using 10-fold cross-validation to select the kernel type (linear, Gaussian and polynomial), regularization parameter and loss parameters (to deal with unbalanced class problem - more negative events than positive ones). Polynomial kernels were selected for both datasets (order 3 for the Greek dataset and 2 for the Spanish). After that, the classifier is dynamically updated on the remaining data (keeping the hyperparameters and kernel type fixed) as described in Section \ref{eventsFiltering}. All the reported results are based on predictions on unseen Twitter events from this remaining data. Overall 375 Twitter events were detected on the Greek dataset and 349 on the Spanish. 

\begin{figure}[!th]
  \centering
\includegraphics[width=0.95\textwidth]{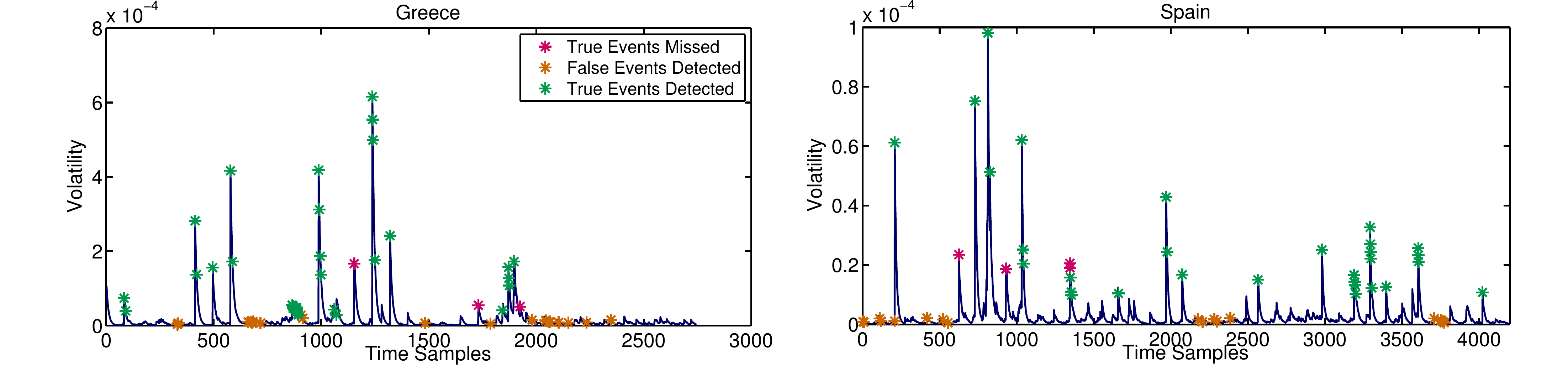}
\caption{Event Detection Results.}
\label{fig:detectedEventsResults}
\end{figure}

We estimate the precision, recall and F1 score of our event detection method by comparing the real label $C_{t_s}(t')$ with the predicted label $L_{t_s}(t')$ for each Twitter event. We create two binary streams $C$ and $L$ with all the real and predicted labels of Twitter events, respectively.
Since there may be stock market events without any matching Twitter event, we update the $C$ and $L$ as follows:\\
\emph{$\forall t' \in \mathcal{T}_{true}$ and $t' \notin \mathcal{U}$, where $\mathcal{U}$ is the set of Twitter event times, create a new label $C_{t'}(t') = 1$ and $L_{t'}(t') = 0$. 
}

In Table \ref{tab:classification} we present the classifier performance for the three different $T_{true}$ thresholds. For both the examined datasets, FED performs better for larger values on $T_{true}$ i.e. when it is trained to detect 'stronger' stock market fluctuations. We also observe slightly improved performance on the Greek dataset. This could be justified considering that during the examined period, Greece was on financial instability which resulted on several stock market jitters.  

\begin{table}[]
\begin{small}
\centering
\begin{tabular}{|l|ll|ll|ll|}
	\hline
  \multirow{ 2}{*}{$\mathbf{T_{true}}$}  & \multicolumn{2} {c|} {\textbf{Precision}}  &  \multicolumn{2} {c|} {\textbf{Recall}}&   \multicolumn{2} {c|}{\textbf{F1}}\\
	& GR  & ES & GR  & ES	& GR  & ES\\
 \hline
 $2 \cdot \langle\mathcal{V'}(s)\rangle_s$ & 0.61 & 0.52 & 0.73 & 0.79 & 0.66 & 0.63\\
 \hline
 $2.5 \cdot \langle\mathcal{V'}(s)\rangle_s$ &  0.64 & 0.62 & 0.84 & 0.69 & 0.72 & 0.66\\
\hline
 $3 \cdot \langle\mathcal{V'}(s)\rangle_s$ &  0.69 & 0.65 & 0.81 & 0.72 & 0.74 & 0.68\\
  \hline
 
\end{tabular}
\caption{Classification Performance }
\label{tab:classification}
\end{small}
\end{table}

In Figure \ref{fig:detectedEventsResults} we present historical volatility of the ATHEX and IBEX stock market indexes, for the months after the training period. We also show the correctly and falsely detected events, as well as the missed events for $T_{true} = 3 \cdot \langle\mathcal{V'}(s)\rangle_s$. According to our results, the proposed mechanism successfully detects most of the stock market jitters purely based on Twitter data. Interestingly, although not specifically trained to do so, all detected stock market events were predicted as positive on Twitter before they appeared on the stock market. Finally there are some Twitter events falsely classified as \emph{positive}. These misclassifications usually occur in bursts. This can be explained by the fact that our approach allows for multiple updated versions of the same event; if one feature vector is misclassified, its subsequent updated versions will be probably misclassified too. One hour after the first misclassified vector occurs, the classifier is updated with the new sample, and consequently avoids repeating the same mistake on any similar subsequent feature vectors.

\subsubsection{Comparison With Baseline Event Detectors}

We compare the performance of our approach with a) a state-of-the-art general-purpose event detector and b) a sentiment-based event detector. For events detected when the stock market is closed, we apply the process used in FED, i.e. such events will be shifted to the opening time of the stock market. If this time is more than 24 hours ahead of the event time (during the weekend), the event will be discarded. If more than one Twitter events are matched to the same stock market time, we keep only the 'stronger' event. The strength of an event is defined based on the event detection method described in the following paragraphs.

\textbf{General-purpose Event Detector.} Although several methods for bursts detection on social media data have been proposed, to the best of our knowledge, this is the first work that attempts to identify events that influence a specific stock market. The most similar approach to FED is EDCoW. Both FED and EDCoW monitor changes on the arrival rates of individual words in order to trigger the detection of an event and they group bursty words based on the correlations among their arrival patterns. However, in contrast to FED, which uses word groups to construct Twitter event features, EDCoW creates a separate event for each word group. Finally, for each event, EDCoW estimates a value $\epsilon$ representing the 'strength' of the event based on the number of its words as well as the correlations among them and filters-out non-significant events (i.e. events with low $\epsilon$ value). We perform event detection in 2-hour windows. Similarly to our approach, we assign a label $C(t_w)$ to each event $E(t_w)$ detected during a window started at time $t_w$ as follows:

\begin{equation}
C(t_w) = 
\left\{
  \begin{array}{rcr}
	1 & \mbox{if $\exists s \in \mathcal{T}_{true}, t_w\leq s \leq t_w + 2h$}\\
	-1 & \mbox{if $\exists s \in \mathcal{T}_{neutral}, t_w\leq s \leq t_w + 2h$}\\
    0 & \mbox{otherwise}\\
    
\end{array}
\right.
\end{equation}

The total number of true positive and false positive events is given by the number of 1- and 0-labels, respectively, in $C$. We also estimate the number of false negative by counting the the number of stock market jitters for which there was no event detected.  In Table \ref{tab:edcowClassification} we present the performance of EDCoW for the three $T_{true}$ thresholds and three different values on the $\gamma$ parameter of EDCoW that is used to define when the correlation between two words (or the autocorrelation of one word) is significant. These results correspond to the optimal threshold on $\epsilon$ value, which is used to filter-out non-significant events (i.e. the threshold for which we achieved the highest F1 score). Note, that such an evaluation favors EDCoW method over FED, as the performance estimates will be positively biased.
In spite of that, the EDCoW performs poorly for all the examined $\gamma$ values. This indicates that it is not feasible to detect Twitter events that influence the stock market solely by searching for bursts in the Twitter stream.

\begin{table}[]
\begin{small}
\centering
\begin{tabular}{|l|l|ll|ll|ll|}
	\hline
 \multirow{ 2}{*}{$\mathbf{\gamma}$} & \multirow{ 2}{*}{$\mathbf{T_{true}}$} & \multicolumn{2} {c|} {\textbf{Precision}}  &  \multicolumn{2} {c|} {\textbf{Recall}}&   \multicolumn{2} {c|}{\textbf{F1}}\\
	& & GR  & ES & GR  & ES	& GR  & ES\\
	\hline
  \multirow{ 3}{*}{10} & $2 \cdot \langle\mathcal{V'}(s)\rangle_s$ & 0.35 & 0.39 & 0.22 & 0.27 & 0.27 & 0.32\\
 &$2.5 \cdot \langle\mathcal{V'}(s)\rangle_s$ & 0.31 & 0.33 & 0.21 & 0.22 & 0.25 & 0.26\\
 &$3 \cdot \langle\mathcal{V'}(s)\rangle_s$ & 0.31 & 0.35 & 0.22& 0.22 & 0.26 & 0.27\\
 	\hline
  \multirow{ 3}{*}{20} & $2 \cdot \langle\mathcal{V'}(s)\rangle_s$ & 0.21 & 0.24 & 0.33 & 0.28 & 0.25 & 0.26\\
 &$2.5 \cdot \langle\mathcal{V'}(s)\rangle_s$  & 0.18 & 0.22 & 0.30 & 0.30 & 0.23 & 0.25\\
 &$3 \cdot \langle\mathcal{V'}(s)\rangle_s$ & 0.18 & 0.23 & 0.31 & 0.32 & 0.23 & 0.27\\
 \hline
  \multirow{ 3}{*}{40} & $2 \cdot \langle\mathcal{V'}(s)\rangle_s$ & 0.19 & 0.25 & 0.60 & 0.47 & 0.29 & 0.33\\
 &$2.5 \cdot \langle\mathcal{V'}(s)\rangle_s$  & 0.17 & 0.22 & 0.56 & 0.49 & 0.26 & 0.30\\
 &$3 \cdot \langle\mathcal{V'}(s)\rangle_s$ & 0.17 & 0.23 & 0.54 & 0.49 & 0.26 & 0.31\\
 \hline
\end{tabular}
\caption{EDCoW Event Detection.}
\label{tab:edcowClassification}
\end{small}
\end{table}

\textbf{Sentiment-based Event Detector.} Given that most studies on the influence of social media on the stock market only examine the impact of text sentiment, we compare FED with a sentiment-based event detector. A direct comparison with existing methods is not feasible, since, to the best of our knowledge, their purpose is either to prove a dependency between social media and stock market or to predict future values rather than the detection of jitters. Thus, we adjust FED in order to use only information about tweets sentiment. In detail, we estimate the weighted sentiment strength index $SSI_W(t)$, described in Section \ref{featuresDescription}, using 2-hour sliding windows with 5 minutes step size. We then apply an event detection method similar to the proposed FED approach: we create an event at time $t$ if $\langle SSI_W(t)\rangle_t \leq SSI_W(t)$ and we update the event when there is a 10\% increase in its sentiment value. We label events as \emph{positive} or \emph{negative} by applying Equation \ref{eq:labelVectors} and we train an Support Vector Machine classifier in order to predict the events classes. In Table \ref{tab:sentimentEventDetection} we present the precision, recall an F1 score of the sentiment-based event detector for the three different $T_{true}$ thresholds. Our results indicate that events classification based solely on sentiment performs poorly, since it is not possible to distinguish between events of negative sentiment which influence stock market (e.g. fears of political instability or bankruptcy) and those that do not (e.g. negative opinions/gossips about politicians).

\begin{table}[]
\begin{small}
\centering
\begin{tabular}{|l|ll|ll|ll|}
 \hline
  \multirow{ 2}{*}{$\mathbf{T_{true}}$}  & \multicolumn{2} {c|} {\textbf{Precision}}  &  \multicolumn{2} {c|} {\textbf{Recall}}&   \multicolumn{2} {c|}{\textbf{F1}}\\
	& GR  & ES & GR  & ES	& GR  & ES\\
 \hline
  $2 \cdot \langle\mathcal{V'}(s)\rangle_s$ & 0.23 & 0.21 & 0.79 & 0.74 & 0.37 & 0.34\\
 \hline
 $2.5 \cdot \langle\mathcal{V'}(s)\rangle_s$ & 0.22 & 0.23 & 0.76 & 0.77 & 0.34 & 0.35\\
 \hline
 $3 \cdot \langle\mathcal{V'}(s)\rangle_s$ & 0.23 & 0.23 & 0.79 & 0.76 & 0.36 & 0.35\\
 \hline
  
\end{tabular}
\caption{Sentiment-based Event Detection.}
\label{tab:sentimentEventDetection}
\end{small}
\end{table}

\begin{figure}[!th]
  \centering
  \includegraphics[width=0.95\textwidth]{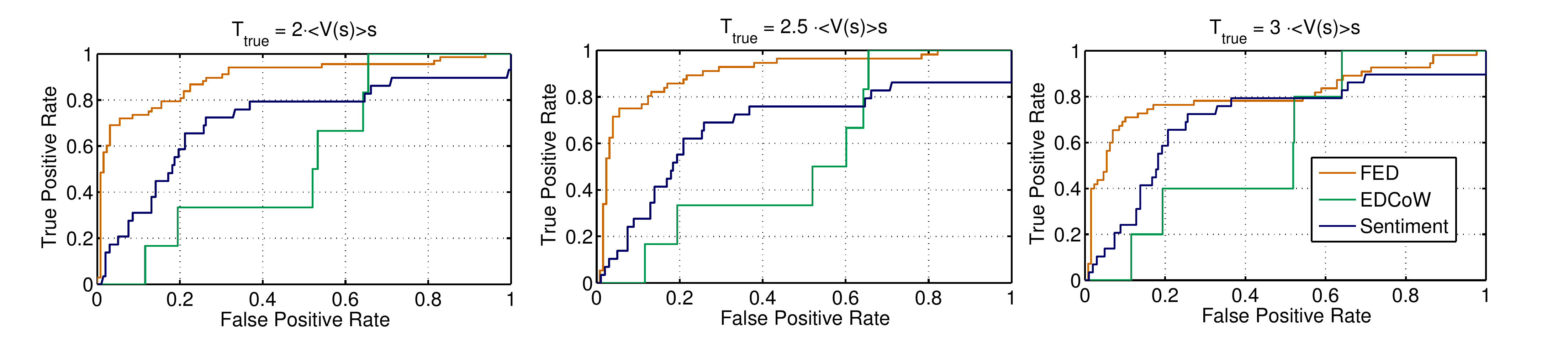}
\caption{ROC Curves.}
\label{fig:rocCurves}
\end{figure}

Finally, in Figure \ref{fig:rocCurves} we present the receiver-operating-curves (roc) for FED, EDCoW and the sentiment-based event detector  (with $\gamma = 40$) for the three $T_{true}$ thresholds. The roc curves for the EDCoW method were created by varying the threshold on $\epsilon$ value used in filtering-out non-significant events.

\subsection{Mutual Information Analysis}
\label{sec:mutualInfo}

In this section we use mutual information to examine the dependence between the Twitter events and the stock market jitters. We represent events in stock market using the binary stream $C$ of real event classes and Twitter events with the binary stream $L$ of predicted Twitter event classes. The binary streams are not i.i.d.. The probability of a stock market jitter will normally be higher when strong fluctuations have been previously observed and lower in more 'stable' periods. Thus, we model the binary streams $C$, $L$ as markov chains by applying the Causal State Splitting Reconstruction (CSSR) algorithm \cite{shalizi2004blind}. CSSR creates a Markov model which best represents the underlying probabilistic model of the streams. The resulted model is a two-state Markov chain (i.e. the probability of having an event labeled as \emph{positive/negative} at time $t$ depends only on the event label at time $t-1$). We denote with $\pi^C_i$, $\pi^L_i$ the probabilities of state $i$ for $C$ and $L$ respectively and with $p^C_{i|j}$, $p^L_{i|j}$ the transition probabilities from state $j$ to state $i$. Then, the entropy rates of $C$, $L$ are estimated as follows \cite{ekroot1993entropy}:

\begin{equation*}
H(C) {=} -\sum_{i=0}^1 \pi^C_i \cdot \sum_{j=0}^1 p^C_{j|i} \log{p^C_{j|i}}
\end{equation*}
\begin{equation*}
H(L) {=} -\sum_{i=0}^1 \pi^L_i \cdot \sum_{j=0}^1 p^L_{j|i} \log{p^L_{j|i}}
\end{equation*}

We measure the reduction of uncertainty about $C$ during a time unit $t$ if we utilize knowledge about $L$ during $t$ by measuring the mutual information rate $MIR(C, L)$ \cite{blanc2011delay} given by the following equation:

\begin{equation}
MIR(C, L) = H(C)+H(L)-H(C, L)
\end{equation}
where $H(C, L)$ denote the joint Shannon entropy of $C$, $L$ estimated as:

\begin{equation}
H(C, L) {=} -\sum_{i=0}^1 \sum_{j=0}^1 \pi^{C,L}_{i,j} {\cdot}
\sum_{k=0}^1 \sum_{l=0}^1 p^{C,L}_{k,l|i,j} \log{ p^{C,L}_{k,l|i,j}}
\end{equation}
where $\pi^{C,L}_{i,j}$ the joint state probability of $C$ and $L$ for states $i$, $j$ respectively and $p^{C,L}_{k,l|i,j}$ the joint transition probability of $C$ and $L$ from states $i$ to $k$ and $j$ to $l$ respectively. 

\begin{figure}[!th]
  \centering
  \includegraphics[width=0.9\textwidth]{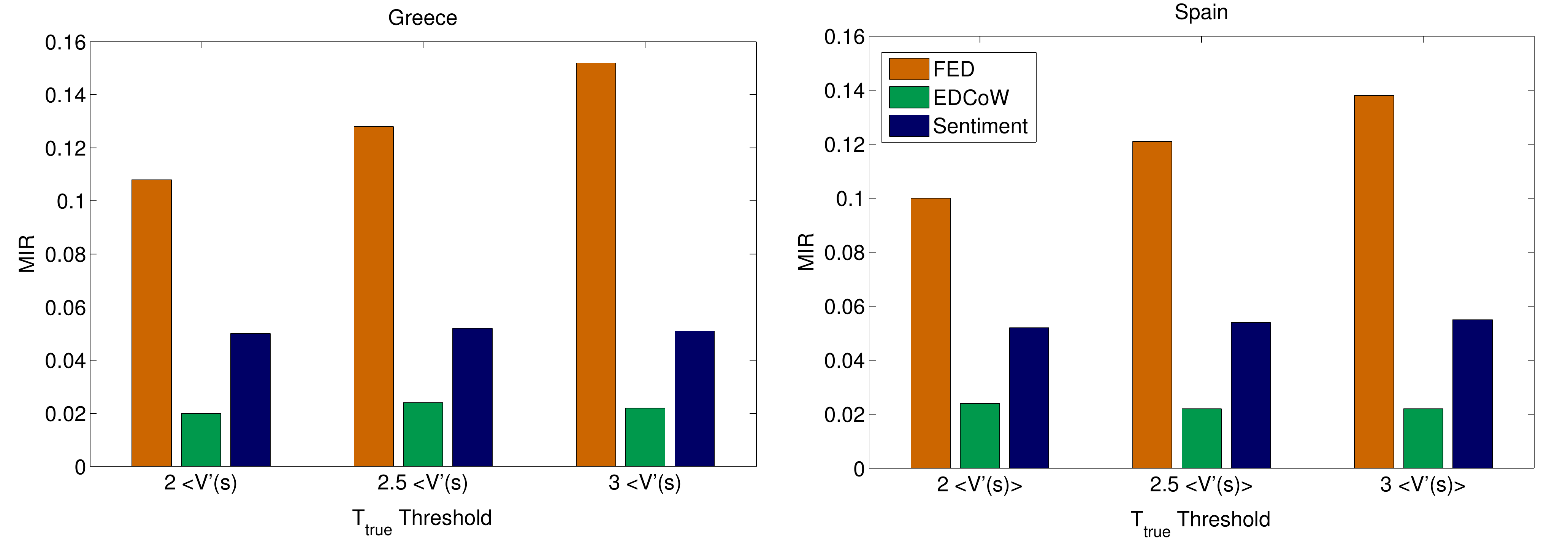}
\caption{Mutual Information Rate between real events and predicted events}
\label{fig:rmi}
\end{figure}

In Figure \ref{fig:rmi} we present the mutual information rate between $C$ and $L$, when $L$ is estimated by applying a) the proposed FED method, b) the EDCoW method and c) the sentiment-based event detector, for the three different $T_{true}$ threshold values. The estimation of $MIR$ is based only on unseen Twitter events (i.e. we do not use the training set). Our results indicate significant dependence between stock market jitters and events detected by the FED approach and much weaker dependence when sentiment-based or EDCoW event detection is applied. 

\section{Conclusion}

\label{conclussions}

In this paper, we present FED (Financial Event Detector), a novel event detection method which focuses on early detection of events in Twitter that influence a specific stock market. We model Twitter data as multi-dimensional feature vectors by utilizing a rich variety of information. We apply feature selection in order to find which of these features are important for distinguishing between events that influence stock market and insignificant events. We demonstrate the benefit of using multiple features for modeling Twitter information, instead of extracting only a sentiment index like previous works, by comparing FED with a sentiment-based event detector. We train a classifier, solely by utilizing stock market data, to recognize which of the detected events will cause strong fluctuations on the examined stock market.  We apply our method to two real-world datasets and we demonstrate that FED achieves up to 74.32\% F1 score. We also show that general-purpose event detectors fail to recognize events that influence stock market. 


\begin{backmatter}

\section*{Competing interests}
  The authors declare that they have no competing interests.

\section*{Author's contributions}
    FT, NB, PT and MM designed the study. FT and NB implemented the method and they collected and analyzed the data. FT prepared the figures. FT, NB, PT and MM wrote the manuscript.

\section*{Acknowledgments}

PT was supported by EPSRC grant no EP/L000296/1 "Personalised Medicine through Learning in the Model Space".
\theendnotes



\newcommand{\BMCxmlcomment}[1]{}

\BMCxmlcomment{

<refgrp>

<bibl id="B1">
  <title><p>Twitter Mood Predicts the Stock Market</p></title>
  <aug>
    <au><snm>Bollen</snm><fnm>J</fnm></au>
    <au><snm>Mao</snm><fnm>H</fnm></au>
    <au><snm>Zeng</snm><fnm>X</fnm></au>
  </aug>
  <source>Journal of Computational Science</source>
  <publisher>Elsevier</publisher>
  <pubdate>2011</pubdate>
  <volume>2</volume>
  <issue>1</issue>
  <fpage>1</fpage>
  <lpage>-8</lpage>
</bibl>

<bibl id="B2">
  <title><p>{Predicting stock market indicators through Twitter {"I hope it is
  not as bad as I fear"}}</p></title>
  <aug>
    <au><snm>Zhang</snm><fnm>X</fnm></au>
    <au><snm>Fuehres</snm><fnm>H</fnm></au>
    <au><snm>Gloor</snm><fnm>PA</fnm></au>
  </aug>
  <source>Proceedings of the 2nd Collaborative Innovation Networks
  Conference</source>
  <pubdate>2011</pubdate>
  <volume>26</volume>
  <fpage>55</fpage>
  <lpage>-62</lpage>
</bibl>

<bibl id="B3">
  <title><p>Predicting financial markets: Comparing survey, news, Twitter and
  search engine data</p></title>
  <aug>
    <au><snm>Mao</snm><fnm>H</fnm></au>
    <au><snm>Counts</snm><fnm>S</fnm></au>
    <au><snm>Bollen</snm><fnm>J</fnm></au>
  </aug>
  <source>arXiv preprint arXiv:1112.1051</source>
  <pubdate>2011</pubdate>
</bibl>

<bibl id="B4">
  <title><p>Emerging topic detection on twitter based on temporal and social
  terms evaluation</p></title>
  <aug>
    <au><snm>Cataldi</snm><fnm>M</fnm></au>
    <au><snm>Di Caro</snm><fnm>L</fnm></au>
    <au><snm>Schifanella</snm><fnm>C</fnm></au>
  </aug>
  <source>Proceedings of the Tenth International Workshop on Multimedia Data
  Mining</source>
  <pubdate>2010</pubdate>
  <fpage>4</fpage>
</bibl>

<bibl id="B5">
  <title><p>Event Detection in Twitter.</p></title>
  <aug>
    <au><snm>Weng</snm><fnm>J</fnm></au>
    <au><snm>Lee</snm><fnm>BS</fnm></au>
  </aug>
  <source>Proceedings of the International Conference on Weblogs and Social
  Media (ICWSM'11)</source>
  <pubdate>2011</pubdate>
  <fpage>401</fpage>
  <lpage>-408</lpage>
</bibl>

<bibl id="B6">
  <title><p>Topicsketch: Real-time bursty topic detection from
  twitter</p></title>
  <aug>
    <au><snm>Xie</snm><fnm>W</fnm></au>
    <au><snm>Zhu</snm><fnm>F</fnm></au>
    <au><snm>Jiang</snm><fnm>J</fnm></au>
    <au><snm>Lim</snm><fnm>EP</fnm></au>
    <au><snm>Wang</snm><fnm>K</fnm></au>
  </aug>
  <source>Proceedings of the 13th International Conference on Data Mining
  (ICDM'13</source>
  <pubdate>2013</pubdate>
  <fpage>837</fpage>
  <lpage>-846</lpage>
</bibl>

<bibl id="B7">
  <title><p>Twevent: segment-based event detection from tweets</p></title>
  <aug>
    <au><snm>Li</snm><fnm>C</fnm></au>
    <au><snm>Sun</snm><fnm>A</fnm></au>
    <au><snm>Datta</snm><fnm>A</fnm></au>
  </aug>
  <source>Proceedings of the 21st ACM international Conference on Information
  and Knowledge Management (CIKM'12)</source>
  <pubdate>2012</pubdate>
  <fpage>155</fpage>
  <lpage>-164</lpage>
</bibl>

<bibl id="B8">
  <title><p>Measuring geographical regularities of crowd behaviors for
  Twitter-based geo-social event detection</p></title>
  <aug>
    <au><snm>Lee</snm><fnm>R</fnm></au>
    <au><snm>Sumiya</snm><fnm>K</fnm></au>
  </aug>
  <source>Proceedings of the 2nd ACM SIGSPATIAL international Workshop on
  Location based Social Networks</source>
  <pubdate>2010</pubdate>
  <fpage>1</fpage>
  <lpage>-10</lpage>
</bibl>

<bibl id="B9">
  <title><p>Earthquake shakes Twitter users: real-time event detection by
  social sensors</p></title>
  <aug>
    <au><snm>Sakaki</snm><fnm>T</fnm></au>
    <au><snm>Okazaki</snm><fnm>M</fnm></au>
    <au><snm>Matsuo</snm><fnm>Y</fnm></au>
  </aug>
  <source>Proceedings of the 19th International Conference on World Wide Web
  (WWW'10)</source>
  <pubdate>2010</pubdate>
  <fpage>851</fpage>
  <lpage>-860</lpage>
</bibl>

<bibl id="B10">
  <title><p>Complex dynamics of our economic life on different scales: insights
  from search engine query data</p></title>
  <aug>
    <au><snm>Preis</snm><fnm>T</fnm></au>
    <au><snm>Reith</snm><fnm>D</fnm></au>
    <au><snm>Stanley</snm><fnm>HE</fnm></au>
  </aug>
  <source>Philosophical Transactions of the Royal Society of London A:
  Mathematical, Physical and Engineering Sciences</source>
  <publisher>The Royal Society</publisher>
  <pubdate>2010</pubdate>
  <volume>368</volume>
  <issue>1933</issue>
  <fpage>5707</fpage>
  <lpage>-5719</lpage>
</bibl>

<bibl id="B11">
  <title><p>Quantifying trading behavior in financial markets using Google
  Trends</p></title>
  <aug>
    <au><snm>Preis</snm><fnm>T</fnm></au>
    <au><snm>Moat</snm><fnm>HS</fnm></au>
    <au><snm>Stanley</snm><fnm>HE</fnm></au>
  </aug>
  <source>Scientific Reports</source>
  <publisher>Nature Publishing Group</publisher>
  <pubdate>2013</pubdate>
  <volume>3</volume>
</bibl>

<bibl id="B12">
  <title><p>Correlating financial time series with micro-blogging
  activity</p></title>
  <aug>
    <au><snm>Ruiz</snm><fnm>EJ</fnm></au>
    <au><snm>Hristidis</snm><fnm>V</fnm></au>
    <au><snm>Castillo</snm><fnm>C</fnm></au>
    <au><snm>Gionis</snm><fnm>A</fnm></au>
    <au><snm>Jaimes</snm><fnm>A</fnm></au>
  </aug>
  <source>Proceedings of the Fifth ACM International Conference on Web Search
  and Data Mining (WSDM'12)</source>
  <pubdate>2012</pubdate>
  <fpage>513</fpage>
  <lpage>-522</lpage>
</bibl>

<bibl id="B13">
  <title><p>Combining technical analysis with sentiment analysis for stock
  price prediction</p></title>
  <aug>
    <au><snm>Deng</snm><fnm>S</fnm></au>
    <au><snm>Mitsubuchi</snm><fnm>T</fnm></au>
    <au><snm>Shioda</snm><fnm>K</fnm></au>
    <au><snm>Shimada</snm><fnm>T</fnm></au>
    <au><snm>Sakurai</snm><fnm>A</fnm></au>
  </aug>
  <source>Proceedings of the 2011 IEEE Ninth International Conference on
  Dependable, Autonomic and Secure Computing (DASC'11)</source>
  <pubdate>2011</pubdate>
  <fpage>800</fpage>
  <lpage>-807</lpage>
</bibl>

<bibl id="B14">
  <title><p>Textual analysis of stock market prediction using breaking
  financial news: The AZFin text system</p></title>
  <aug>
    <au><snm>Schumaker</snm><fnm>RP</fnm></au>
    <au><snm>Chen</snm><fnm>H</fnm></au>
  </aug>
  <source>ACM Transactions on Information Systems (TOIS)</source>
  <publisher>ACM</publisher>
  <pubdate>2009</pubdate>
  <volume>27</volume>
  <issue>2</issue>
  <fpage>12</fpage>
</bibl>

<bibl id="B15">
  <title><p>An Experiment in Integrating Sentiment Features for Tech Stock
  Prediction in Twitter</p></title>
  <aug>
    <au><snm>Vu</snm><fnm>TT</fnm></au>
    <au><snm>Chang</snm><fnm>S</fnm></au>
    <au><snm>Ha</snm><fnm>QT</fnm></au>
    <au><snm>Collier</snm><fnm>N</fnm></au>
  </aug>
  <source>Proceedings of the 24th International Conference on Computational
  Linguistics</source>
  <pubdate>2012</pubdate>
  <fpage>23</fpage>
</bibl>

<bibl id="B16">
  <title><p>Heart and soul: Sentiment strength detection in the social web with
  sentistrength</p></title>
  <aug>
    <au><snm>Thelwall</snm><fnm>M</fnm></au>
  </aug>
  <source>Proceedings of the CyberEmotions</source>
  <publisher>Citeseer</publisher>
  <pubdate>2013</pubdate>
  <fpage>1</fpage>
  <lpage>-14</lpage>
</bibl>

<bibl id="B17">
  <title><p>Automatic keyword extraction from individual documents</p></title>
  <aug>
    <au><snm>Rose</snm><fnm>S</fnm></au>
    <au><snm>Engel</snm><fnm>D</fnm></au>
    <au><snm>Cramer</snm><fnm>N</fnm></au>
    <au><snm>Cowley</snm><fnm>W</fnm></au>
  </aug>
  <source>Text Mining</source>
  <pubdate>2010</pubdate>
  <fpage>1</fpage>
  <lpage>-20</lpage>
</bibl>

<bibl id="B18">
  <title><p>Weka: Practical machine learning tools and techniques with Java
  implementations</p></title>
  <aug>
    <au><snm>Witten</snm><fnm>IH</fnm></au>
    <au><snm>Frank</snm><fnm>E</fnm></au>
    <au><snm>Trigg</snm><fnm>LE</fnm></au>
    <au><snm>Hall</snm><fnm>MA</fnm></au>
    <au><snm>Holmes</snm><fnm>G</fnm></au>
    <au><snm>Cunningham</snm><fnm>SJ</fnm></au>
  </aug>
  <pubdate>1999</pubdate>
</bibl>

<bibl id="B19">
  <title><p>Feature subset selection: a correlation based filter
  approach</p></title>
  <aug>
    <au><snm>Hall</snm><fnm>MA</fnm></au>
    <au><snm>Smith</snm><fnm>LA</fnm></au>
  </aug>
  <source>Proceedings of the International Conference on Neural Information
  Processing and Intelligent Information Systems</source>
  <pubdate>1997</pubdate>
  <fpage>855</fpage>
  <lpage>-858</lpage>
</bibl>

<bibl id="B20">
  <title><p>A comparative study on feature selection in text
  categorization</p></title>
  <aug>
    <au><snm>Yang</snm><fnm>Y</fnm></au>
    <au><snm>Pedersen</snm><fnm>JO</fnm></au>
  </aug>
  <source>ICML</source>
  <pubdate>1997</pubdate>
  <volume>97</volume>
  <fpage>412</fpage>
  <lpage>-420</lpage>
</bibl>

<bibl id="B21">
  <title><p>Blind construction of optimal nonlinear recursive predictors for
  discrete sequences</p></title>
  <aug>
    <au><snm>Shalizi</snm><fnm>CR</fnm></au>
    <au><snm>Shalizi</snm><fnm>KL</fnm></au>
  </aug>
  <source>Proceedings of the 20th Conference on Uncertainty in Artificial
  Intelligence</source>
  <pubdate>2004</pubdate>
  <fpage>504</fpage>
  <lpage>-511</lpage>
</bibl>

<bibl id="B22">
  <title><p>The entropy of Markov trajectories</p></title>
  <aug>
    <au><snm>Ekroot</snm><fnm>L</fnm></au>
    <au><snm>Cover</snm><fnm>TM</fnm></au>
  </aug>
  <source>IEEE Transactions on Information Theory</source>
  <publisher>IEEE</publisher>
  <pubdate>1993</pubdate>
  <volume>39</volume>
  <issue>4</issue>
  <fpage>1418</fpage>
  <lpage>-1421</lpage>
</bibl>

<bibl id="B23">
  <title><p>Delay independence of mutual-information rate of two symbolic
  sequences</p></title>
  <aug>
    <au><snm>Blanc</snm><fnm>JL</fnm></au>
    <au><snm>Pezard</snm><fnm>L</fnm></au>
    <au><snm>Lesne</snm><fnm>A</fnm></au>
  </aug>
  <source>Physical Review E</source>
  <publisher>APS</publisher>
  <pubdate>2011</pubdate>
  <volume>84</volume>
  <issue>3</issue>
  <fpage>036214</fpage>
</bibl>

</refgrp>
} 

\end{backmatter}

\end{document}